\documentclass[12pt,preprint]{aastex}

\usepackage{times}

\newcommand{\msun}{M_\odot}

\catcode`\@=11
\newcommand{\gapprox}{\mathrel{\mathpalette\@versim>}}
\newcommand{\lapprox}{\mathrel{\mathpalette\@versim<}}
\newcommand{\propapprox}{\mathrel{\mathpalette\@versim\propto}}
\newcommand{\@versim}[2]
  {\lower3.1truept\vbox{\baselineskip0pt\lineskip0.5truept
\ialign{$\m@th#1\hfil##\hfil$\crcr#2\crcr\sim\crcr}}}
\catcode`\@=12

\shorttitle{An X-ray Analysis of SNR G284}
\shortauthors{WILLIAMS ET AL.}

\begin{document}

\title{Magnesium-rich Ejecta in the SNR G284.3$-$1.8 Around the High-Mass Gamma-Ray Binary 1FGL J1018.6$-$5856}

\author{Brian J. Williams,\altaffilmark{1}
Blagoy Rangelov,\altaffilmark{2}
Oleg Kargaltsev,\altaffilmark{2}
George G. Pavlov,\altaffilmark{3}
}

\altaffiltext{1}{CRESST and X-ray Astrophysics Laboratory, NASA/GSFC, 8800 Greenbelt Road, Greenbelt, MD, Code 662, brian.j.williams@nasa.gov}
\altaffiltext{2}{Department of Physics, The George Washington University, Washington, DC 20052}
\altaffiltext{3}{Pennsylvania State University, 525 Davey Laboratory, University Park, PA 16802}

\begin{abstract}

We present results from two {\it Chandra} observations of the 16.6 d
X-ray/$\gamma$-ray high-mass binary 1FGL~J1018.6$-$5856 located at the
center of the supernova remnant G284.3$-$1.8. The binary spectra,
separated by 0.25 in binary phase, are fit with an absorbed power-law
model with $\Gamma\approx1.7-1.8$ for both observations (the flux
during the second observation is a factor of 1.7 smaller).  In the
high-resolution ACIS-I image we found a hint of extended emission
$\approx2''$--$3''$ southeast of the binary, significant at the
$3\sigma$ level. Binary evolution codes reproduce the system's
observed properties with two massive stars with an initial 18-day
period, undergoing mass transfer and leaving behind a heavy $\approx2$
$\msun$ neutron star. The initial mass of the progenitor star in this
scenario is 27 $\pm 4$ $\msun$. \emph{Chandra} and {\it XMM-Newton}
images of the remnant show it has a relatively low X-ray surface
brightness. The two brightest regions of extended X-ray emission, with
luminosities $\sim10^{33}$~erg~s$^{-1}$ for $d=5$~kpc, lie in the
northern and western portions and show significantly different
spectra. The northern patch is consistent with shocked ISM, with a low
temperature and long ionization timescale. However, the western patch
is dominated by ejecta, and shows significantly enhanced Mg content
relative to other ejecta products. The abundance ratios inferred
resemble those from the Large Magellanic Cloud remnant N49B. To our
knowledge, this is only the second case of such Mg-rich ejecta found
in an SNR. Nucleosynthesis models for core-collapse SNe predict
Mg-rich ejecta from very massive progenitors of $>25$ $\msun$.

\end{abstract}


\section{Introduction}

The ejecta from supernova (SN) explosions enrich the interstellar
medium of the Galaxy with heavy elements over thousands of cubic
parsecs. Identification of the ejecta products of SNe is most easily
done via analysis of the X-ray emission from young to middle-aged
supernova remnants (SNRs), after the reverse shock has propagated back
into the ejecta and heated it to X-ray emitting temperatures. One of
the biggest open questions in SN research is what types of stellar
progenitor systems lead to what types of SNe and SNRs. In only a
handful of cases have astronomers directly observed the pre-SN star in
archival images, and this situation is unlikely to change in the near
future. The fundamental problem is that SNe are plentiful but distant
and point-like, while SNRs are local and resolvable but limited in
number.

SNR G284.3$-$1.8 (also known as MSH 10-53; hereafter G284) has been
identified through radio \citep{milne89}, optical \citep{ruiz86}, and
X-ray \citep{abramowski12} observations, and is perhaps $10^4$~years
old \citep{ruiz86} with a distance of 5.4$^{+4.6}_{-2.1}$ kpc
\citep{2011PASP..123.1262N}. It is interacting with a molecular cloud
in the north. An analysis of the extended X-ray emission seen in G284
was first reported by \citet{abramowski12}, who noted that X-ray
emission was present in the northern portion of the remnant, just
behind a filament identifiable in the radio and optical images.

At the center of the remnant lies the bright X-ray point source
1FGL~J1018.6$-$5856 (hereafter J1018). Identified as a high-mass
binary by \citet{2011ATel.3221....1C}, J1018 reveals a 16.6~d periodic
modulation in the \emph{Fermi} LAT observations
\citep{2012Sci...335..189F}. This is one of the few binaries seen in
$\gamma$-rays \citep{2014AN....335..301K}. \citet{2011ATel.3228....1P}
reported results from the first \emph{Chandra} and \emph{XMM-Newton}
observations of J1018 around binary phases $\phi=0.3$ and $\phi=0.6$,
respectively (zero phase corresponds to the maximum of the $1-10$~GeV
flux, MJD~$55403.3\pm0.4$), and also found the high-mass companion in
2MASS data. {\emph{Suzaku} \citep{2014efxu.conf..244T} and
  \emph{Swift} \citep{2013ApJ...775..135A,2015arXiv150502096A}
  observed J1018 covering the entire orbital period, confirming the
  periodicity seen in $\gamma$-rays. The most recent period
  measurement, $16.544\pm0.008$~d, was provided by
  \citet{2015arXiv150502096A}, who found a large X-ray flux increase
  at $\phi=0$ and an additional maximum at $\phi=0.3-0.4$. J1018
  exhibits a correlation between X-ray flux and photon index, common
  to several $\gamma$-ray binaries.}

J1018 is one of two high-mass $\gamma$-ray binaries seen inside of an
SNR (along with SS~433 in W50), offering a rare opportunity to study
the remains of the explosion that would produce such a system. The
spectral type of the companion star places tighter constraints on the
exploded star than would otherwise be possible. Ground-based
observations revealed a massive star of spectral type O6V((f))
($V\approx13.5$; \citealt{2011PASP..123.1262N}) and mass of
$\sim30$~$\msun$, very similar to that of the $\gamma$-ray binary LS
5039. J1018 was observed with the H.E.S.S. observatory by
\citet{2015arXiv150302711H}, who report a $9\sigma$ detection of the
source HESS~J1018$-$589A (spatially coincident with J1018). The
detected $\gamma$-ray variability, both in GeV and TeV, confirms the
association between the H.E.S.S. source and
J1018. \citet{2015arXiv150302711H}, based on the shape of the TeV
phaseogram and source spectrum, suggest that J1018 is a
low-inclination, low-eccentricity system.

In this {\em Letter}, we report our analysis of the X-ray emission
from J1018 and G284. We use newly-obtained {\it Chandra} data,
supplemented with archival {\it XMM-Newton} data, to examine the
extended X-ray emission from the SNR and do spectral and spatial
analysis of the binary.

\section{Observations and Data Reduction}

We use two sets of \emph{Chandra} observations (Table~1), totaling
72~ks, taken with the ACIS-I instrument in timed exposure mode and
telemetered in ``very faint'' format. We processed the data using the
\emph{Chandra} Interactive Analysis of Observations
(CIAO\footnote{\url{http://cxc.harvard.edu/ciao/index.html}}) software
(ver. 4.6) and \emph{Chandra} Calibration Data Base (CALDB),
ver. 4.6.3. Spectra were analyzed using {\it XSPEC} (ver. 12.8.2),
which contains the latest atomic data from {\it AtomDB} 3.0.

We simulate the \emph{Chandra} PSF using
ChaRT\footnote{http://cxc.harvard.edu/chart/} (Chandra Ray Trace) to
study the morphology of the emission in the immediate vicinity of
J1018. The ChaRT output was supplied to MARX
v.5.1\footnote{http://space.mit.edu/CXC/marx/} software to obtain the
final image. We used a blurring (AspectBlur parameter in MARX) value
of $0\farcs19$, as recommended by the {\it Chandra} Help Desk.

\begin{deluxetable}{lccr}
\tablecaption{List of {\it Chandra} observations}
%\tabletypesize{\scriptsize}
\tablewidth{0pt}
\tablehead{
\colhead{ObsId} & \colhead{Date} & \colhead{$\phi$\tablenotemark{a}} & \colhead{Exp\tablenotemark{b}}}
\startdata
14657 & 2013-12-24 & $0.424\pm0.044$ & 44.40(0.031) \\
16560 & 2013-12-29 & $0.671\pm0.044$ & 27.68(0.019) \\
\enddata
\tablenotetext{a}{Orbital phase of the middle of the observational interval. Zero phase corresponds to the $\gamma$-ray peak, MJD $55403.3\pm0.04$; the binary period is $16.544\pm0.008$~d \citep{2015arXiv150502096A}}.
\tablenotetext{b}{Exposure in units of ks, and the corresponding phase interval (in parentheses).}
\end{deluxetable}

For our analysis of the G284 emission, discussed below in
Section~\ref{snr}, we supplement our {\it Chandra} observations with
archival {\it XMM-Newton} observations to increase the signal-to-noise
ratio. G284 was observed by {\it XMM-Newton} on 2013 Jan 9 for a total
of 105 ks (ObsID 0694390101; PI: De Luca). EPIC-MOS observations were
taken in Full Frame mode with a medium filter, while EPIC-pn data were
taken in Small Window mode with a thin filter. The pn data captures
only emission from the binary (analyzed by
\citealt{2015arXiv150502096A}), not the extended emission. We
processed the MOS data with XMM Science Analysis Software (SAS)
ver. 13.5. To filter the data for flares, we used the {\it mos-filter}
routine as part of the ESAS software, available within SAS. After time
filtering, approximately 79 and 91 ks of ``good'' data remained for
the MOS 1 and 2 detectors, respectively.

Guided by the {\it Chandra} images, we selected two regions of
extended emission to analyze, labeled ``north'' and ``west.'' Both are
shown in Figure~\ref{images}. These regions have areas of $\approx
10.9$ and $\approx 16.2$ arcmin$^{2}$, respectively. Background
subtraction for these two regions is difficult due to the large size
of the remnant compared with the {\it Chandra} field of view. However,
based on a comparison of the {\it Chandra} and {\it XMM} images, we
found off-source regions that appear on both the ACIS-I and MOS arrays
and do not appear to have any associated emission from the SNR.

\section{Results \& Discussion}
\label{disc}

\subsection{1FGL~J1018.6$-$5856}

\subsubsection{Spectral and Imaging Analysis}

We extracted spectra for J1018 from the two \emph{Chandra}
observations. The spectra suffer from moderate pile-up (3,294 and
1,498 counts in the $r=3\farcs2$ aperture correspond to 0.24 and 0.17
counts per frame in ObsID 14657 and 16560, respectively). To correct
for pile-up, we used the Davis (2001) model with the grade migration
parameter $\alpha=0.5$. Fits with an absorbed power-law model (with
the XSPEC \texttt{phabs} model) give hydrogen column densities $N_{\rm
  H}=9.0\pm0.9$ and ($9.2\pm1.1)\times10^{21}$~cm$^{-2}$; photon
indices $\Gamma=1.71\pm0.12$ and $1.77\pm0.17$; absorbed (unabsorbed)
fluxes $F_{\rm 0.5-8~keV}=1.5\pm0.1$ ($2.3\pm0.2$) and $0.9\pm0.1$
$(1.4\pm0.2)\times10^{-12}$~erg~s$^{-1}$~cm$^{-2}$; and reduced
$\chi^2=1.07$ and 0.75 for ObsIDs 14657 and 16560, respectively. All
uncertainties are $1\sigma$. The derived $N_{\rm H}$ and $\Gamma$ do
not change significantly between the individual fits. The flux in the
second observation is a factor of 1.7 smaller than the first one, in
accordance with the orbital phase dependance of the flux.
 
Comparing our fit for the first {\sl Chandra} observation with the An
et al.\ (2015) fit for the {\sl XMM-Newton} observation taken at a
close binary phase ($\phi_{\rm CXO}=0.42$, $\phi_{\rm XMM}=0.33$), we
find that our $N_{\rm H}$ is somewhat larger and the slope is somewhat
steeper, but the values are consistent at the $3\sigma$ level. Because
of the lack of pile-up, the {\sl XMM-Newton} measurements are probably
more accurate.

\begin{figure}
\includegraphics[width=7cm]{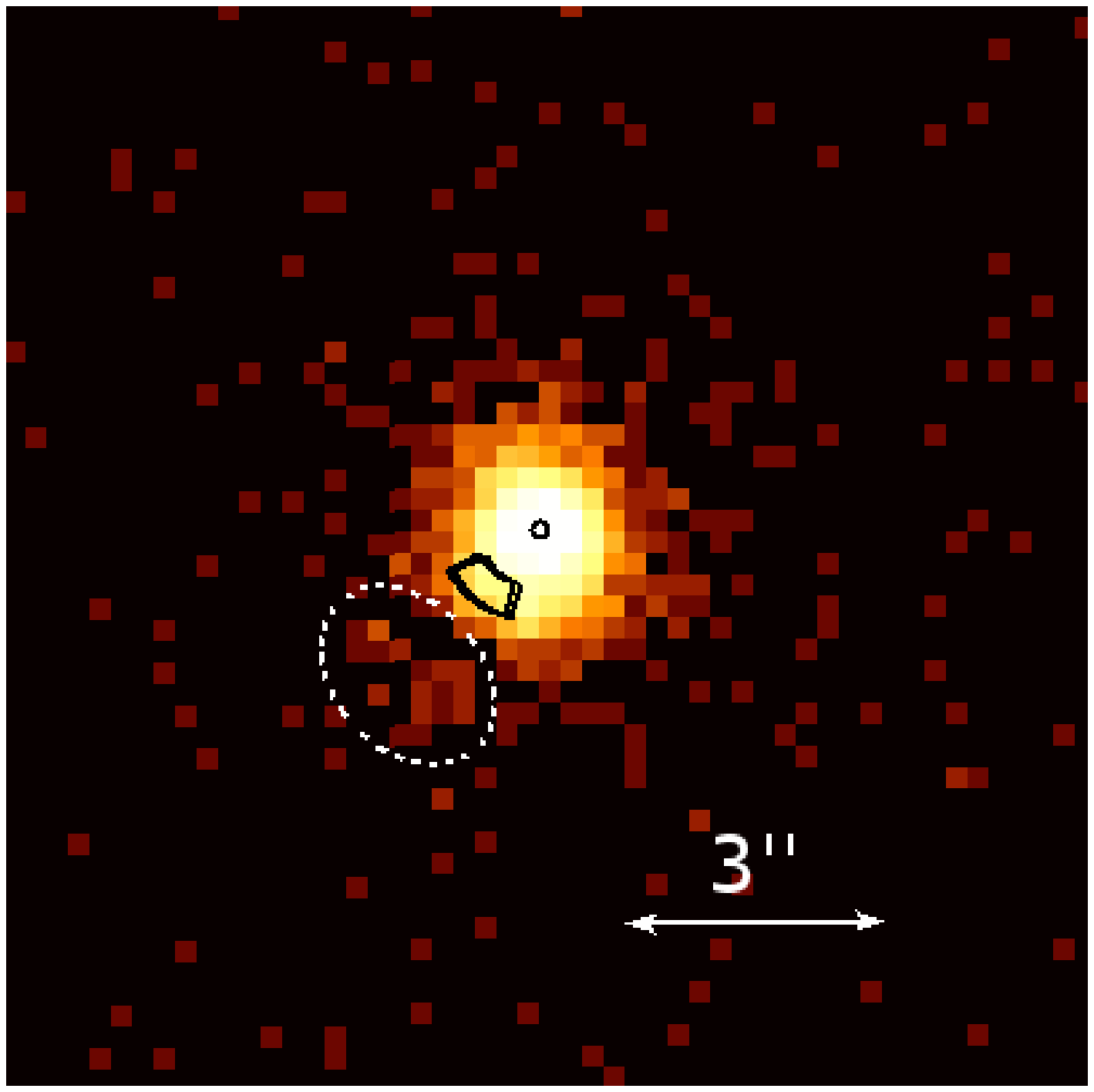}
\includegraphics[width=9.5cm]{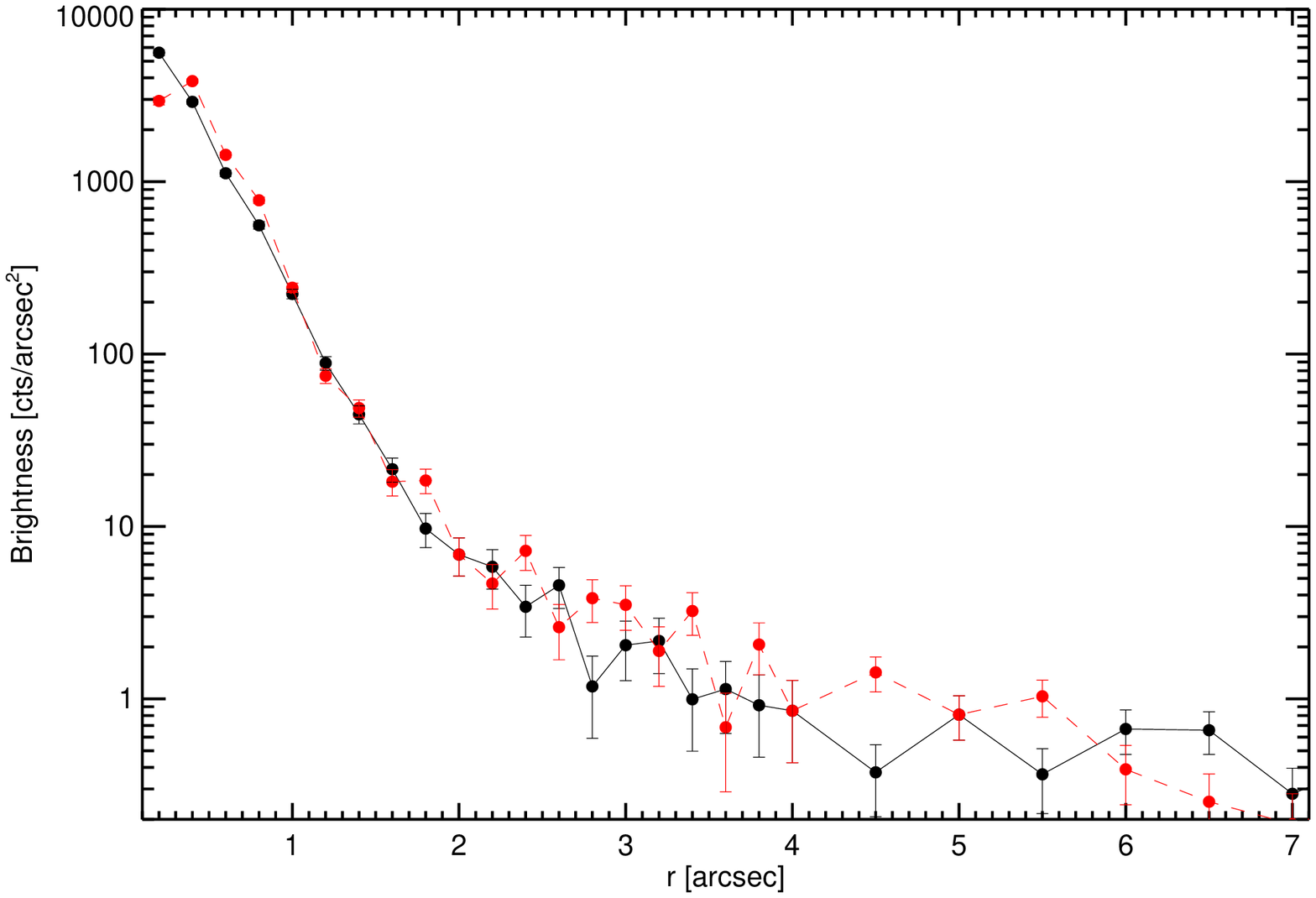}
\caption{\emph{Left:} ACIS-I merged image of J1018 (pixel size
  $0\farcs25$). Small black circle shows the best-fit centroid
  position. The black sectors show the location of the known {\sl
    Chandra} mirror artifact. A slight excess is seen within the white
  ellipse, see text for details. North is up, East is to the
  left. \emph{Right:} Observed surface brightness radial profile of J1018 (black) and modeled PSF (red) with account for pile-up, see Section 3.1.1  for details. 
\label{surfb}}
\end{figure}

\begin{figure}
\includegraphics[width=16cm]{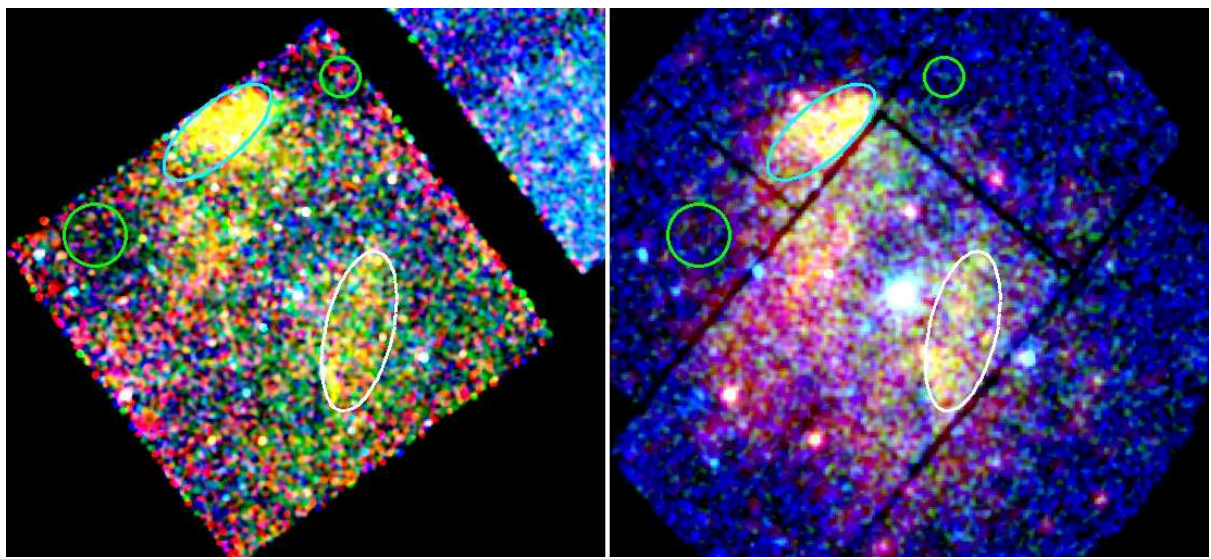}
\caption{{\it Left:} {\it Chandra} image of G284, with 0.5-1.2
  keV emission in red, 1.2-2.0 keV in green, and 2.0-7.0 keV in
  blue. {\it Right:} {\it XMM-Newton} EPIC-MOS 2 image of the
  remnant, with identical color bands. Overlaid are the extraction
  regions, with the ``north'' region in cyan, the ``west'' in white,
  and the backgrounds in green.
\label{images}
}
\end{figure}

We also investigate the spatial morphology of J1018 in the ACIS-I
images and compare the surface brightness radial profile with a
modeled PSF\footnote{We used MARX 5.1 with $AspecBlur=0.19$ and
  pile-up parameter $\alpha=0.5$.} (Figure~\ref{surfb}). Overall, the
observed profile matches the simulated PSF. There appears to be a
mismatch between the simulated profile and the observed profile at
$r<1''$, but this is likely due to an inacurate pile-up modeling in
MARX\footnote{http://space.mit.edu/ASC/MARX/indetail/pileup.html}.  In
addition, the PSF model is not very reliable at small distances due to
the presence of the asymmetric PSF
anomaly\footnote{http://cxc.harvard.edu/proposer/POG/html/chap4.html}
around $r=0\farcs8$, which is not simulated by MARX. Including the
pile-up correction in the MARX 5.1 simulation does not change the
profile substantially beyond $1''$, which is expected for moderate
pileup.

Analyzing the high-resolution images of J1018, we found an enhancement
(24 photons in 0.5--8 keV within the ellipse shown in Figure~1, for
the merged image) located $\approx2\farcs2$ away from the best-fit
centroid of the source at a position angle (PA) of $\approx
140^{\circ}$ (counted East of North).  The enhancement, visible in
both observations, has $\approx 3\sigma$ significance estimated from
comparison with the background (which includes PSF wings) measured in
the $1\farcs75<r<3''$ annulus for the merged image. The expected
enhancement due to the known mirror
artifact\footnote{http://cxc.harvard.edu/ciao/caveats/psf\_artifact.html}
has a very similar PA, but is located significantly closer to the
source center ($r\approx0\farcs8$) and is unlikely the cause of the
enhancement. Its appearance is similar to the fast-moving extended
feature we recently discovered in the vicinity of B1259--63
$\gamma$-ray binary (Pavlov et al.\ 2015). Deeper followup
\emph{Chandra} observations are needed to study this extended emission
and its tentative connection to J1018.

\subsubsection{Modeling of Binary Evolution}

Although the optical counterpart of J1018 has been found and the
binary nature of this system is firmly established, it remains unknown
whether the compact object is a neutron star (NS) or a black hole
(BH). We use the known binary properties (orbital period and mass of
the donor star) to place constraints on the compact object. While
there is still some uncertainty about the exact range of masses of
progenitors that end their lives as NSs and those that become BHs,
most models predict that, for solar or lower metallicities, the
transition between NSs and BHs occurs for stars with initial masses
somewhere in the range of $\sim18-25$~M$_\odot$ (e.g.,
\citealt{1999ApJ...522..413F,heger03}). For metallicities much larger
than solar, stellar evolution models predict that massive stars
develop substantial winds, which cause enough mass loss that the end
product is a NS rather than a BH \citep{heger03}. For our simulation
we assume solar metallicity as the system is in the Galactic disk. We
first use the single star evolution (SSE) code
\citep{2000MNRAS.315..543H} to study the evolution of stars with solar
metallicity. We find that stars with initial mass of ~20 M$_\odot$ and
higher would end up as BHs, in agreement with the above. If J1018
started and evolved as a detached system (no mass transfer ever
happens), the progenitor star must have burned its hydrogen faster
than the $30M_{\odot}$ star and hence it should have been even more
massive. Therefore, in this scenario the compact object should be a
BH. However, the lack of mass transfer would require a much larger
than the currently observed ($16.6$~d) initial orbital period, which
could only increase (since the binary is assumed wide and there is no
interaction to cause spiraling in) after the progenitor explodes as
supernova. Thus we conclude that the system must have started as a
tight binary and undergone mass transfer.

However, evolution of massive stars in tight binaries is more
complicated because mass transfer can proceed from one star to the
other, rendering the SSE code inapplicable. Therefore, we used binary
stellar evolution (BSE) models \citep{2002MNRAS.329..897H}. We ran a
grid of 80,000 simulations with the following parameters:
$M_1=10-35$~M$_\odot$ (current donor star), $M_2=12-50$~M$_\odot$
(SN/compact object progenitor with initial mass always greater than
M1), binary period $P=5-50$~d, eccentricity $e=0-0.9$. We find 115
systems that end up with $P=16.6\pm1.5$~d, and $M_1=31\pm3$~M$_\odot$
right after the SN explosion of star 2. A large fraction of these
simulated binaries are disrupted immediately after the SN, but we are
not interested in such systems. Our findings show that in order to
reproduce the observed binary parameters we need the following initial
conditions: $M_1^{\rm ini}=13.4\pm2.5$~M$_\odot$, $M_2^{\rm
  ini}=26.7\pm3.7$~M$_\odot$, $P=18\pm11$~d. The simulations resulted
in a heavy ($2.2\pm0.4$) NS as the compact object in 95\% of the
cases, while the O6V donor star gained mass (from $13.4$~M$_\odot$ to
31~M$_\odot$) via mass transfer. Note that the systems that match the
above criteria have high eccentricity ($e=0.57\pm0.24$), while
\citet{2015arXiv150302711H} claim (based on the TeV observations) that
the eccentricity of J1018 should be closer to 0.

\subsection{SNR G284.3$-$1.8}
\label{snr}

The northern region analyzed in \citet{abramowski12} roughly
corresponds to our ``north'' region. We extend their analysis,
supplementing it with additional {\it Chandra} and deeper {\it
  XMM-Newton} data (their analysis used a 20 ks {\it XMM}
observation), while expanding the analysis to our ``west'' region as
well. While we do not show it here, we have also found that the
northern filament, identifiable in radio, H$\alpha$, and X-rays, is
present in the WISE 22 $\mu$m IR data. IR emission from the forward
shock in SNRs is common, even in more evolved SNRs \citep{sankrit14},
and generally results from dust heated in the post-shock environment.

We extracted spectra from the north region using only the MOS 2 and
{\it Chandra} observations, since this region falls on one of the dead
MOS 1 chips. For the west region, data from MOS 1, MOS 2, and {\it
  Chandra} are used. After background subtraction, slight residuals
remained in the MOS 1 \& 2 spectra from instrumental lines of Al and
Si K$\alpha$ at 1.49 and 1.75 keV, respectively. We excised small
regions around both of these lines (1.45-1.53 keV and 1.71-1.79 keV)
from the MOS 1 \& 2 spectra, but kept those excised regions in the
{\it Chandra} spectra. For each region, we perform joint fits to all
available data, with all parameters tied together. The column density,
is taken from the fits to the X-ray spectrum of the point source, and
is frozen in the fits to 8$\times 10^{21}$ cm$^{-2}$, the same value
as \citet{abramowski12} used. We binned the data to a minimum of 50
counts per spectral bin for all SNR datasets.

\begin{figure}
\includegraphics[width=8cm]{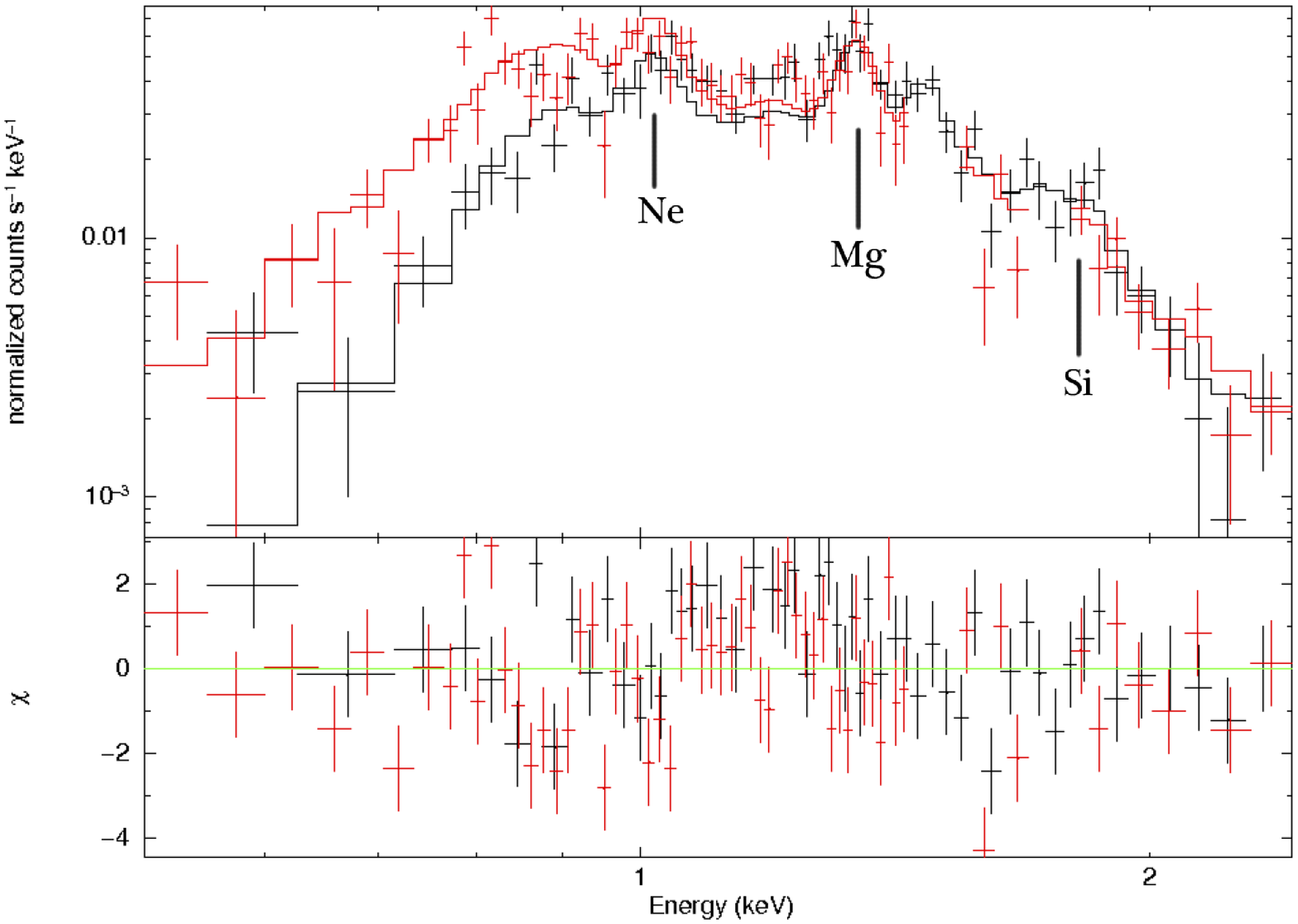}
\includegraphics[width=8cm]{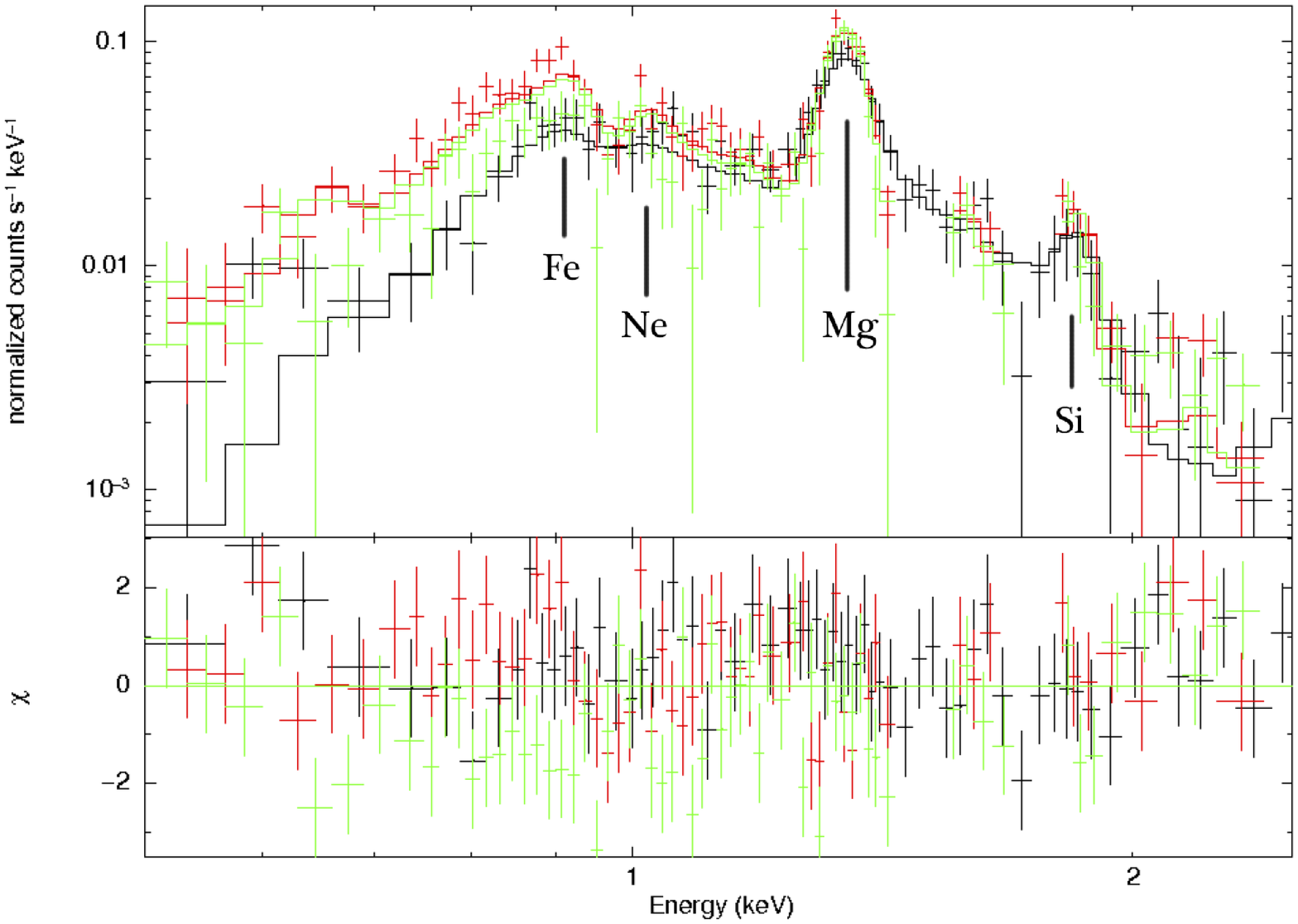}
\caption{{\it Left:} X-ray spectra from the ``north'' region of G284,
  with {\it Chandra} ACIS-I data shown in black and {\it XMM-Newton}
  MOS 2 data shown in red. A fit from a single absorbed {\it vpshock}
  model is overlaid. Fits were performed jointly to both data sets
  simultaneously using {\it XSPEC}. {\it Right:} X-ray spectra from
  the ``west'' region of G284, with {\it Chandra} ACIS-I data shown in
  black, and {\it XMM-Newton} MOS 1 \& 2 shown in red and green,
  respectively. Overlaid is an absorbed {\it vpshock} model fit.
\label{spectra}
}
\end{figure}

We show the spectra from the north and west regions in
Figure~\ref{spectra}. We fit these spectra with an absorbed plane
shock model with variable abundances (\texttt {phabs*vpshock}),
assuming cosmic abundances from \citet{wilms00}. The temperature and
ionization timescale ($\tau$ $\equiv$ $\int^{t}_{0}$ $n_{e}\ dt$) were
allowed to float freely, as were abundances of Ne, Mg, and Fe. Si was
free as well, but was tied together with S due to the poor statistics
of the S line. We kept O frozen at the solar value for two reasons:
with the fairly high absorption towards the remnant, O lines are only
barely visible, if present at all, and we wanted the abundances of the
other elements to be normalized to some fiducial value (the abundance
of O).

For the north region, we find normal cosmic abundances for Ne and Mg,
with subsolar abundances for Si ($+$S) and Fe. Subsolar abundances are
often seen in the X-ray spectra of middle-aged SNRs such as the Cygnus
Loop \citep{tsunemi07}. The relatively long ionization timescale and
lower temperature leads us to concur with the analysis of
\citet{abramowski12} that this emission is dominated by shocked
ISM. The ionization timescale is fit as $4.6\times
10^{12}$~cm$^{-3}$~s, but is unbound on the upper end (lower-bound of
1.9 $\times 10^{12}$). An ionization timescale this long makes the
spectrum almost indistinguishable from a plasma in collisional
ionization equilibrium, again, not unexpected for the shocked ISM from
an older remnant.

The spectra of the west region show an entirely different emission
profile.  They are relatively well-fit (reduced $\chi^{2}$ of 1.4 for
184 d.o.f.) by the same absorbed plane-shock model with variable
abundances with the parameters shown in
Table~\ref{params}. Immediately obvious is the strong Mg line at 1.35
keV. Ne and Fe are both consistent within errors with solar
abundances, while Si (and S) may be slightly overabundant. However,
the overabundance of Mg is about a factor of 4.5 higher with respect
to O. This overabundance, combined with the higher temperature ($\sim
1$ keV) and much lower ionization timescale ($\sim 1 \times 10^{11}$
cm$^{-3}$ s) implies that the emission in this region is dominated by
Mg-rich ejecta, heated by the reverse shock. This conclusion is not
sensitive to the choice of model; a regular NEI model with variable
abundances (model \texttt {vnei} in XSpec) returns similar values.

The spectral fits for this region are similar to those reported in
\citet{park03} for the LMC remnant N49B. Their analysis is similar to
ours, albeit with LMC abundances, different from those in the
Galaxy. They find an enrichment of Mg and Si (though their Si
abundance is higher than ours), while Ne and Fe are not enhanced. They
measure a Mg/O ratio of 4.2 from a small portion in the interior of
N49B, within errors of our measurement of 4.5 in G284. As pointed out
by \citet{park03}, the high Mg/O ratio is surprising, as
nucleosynthetic models predict that significant quantities of Mg
produced in core-collapse SNe should be accompanied by significant
quantities of O \citep{thielemann96}. While examples of significant O
without Mg are plentiful, it is unexpected to find significant Mg
without O.

As \citet{park03} show, deriving a total mass of Mg ejected by the SN
from X-ray spectral analysis is very difficult, with estimates varying
by more than an order of magnitude depending on the assumptions
made. Such an estimate is beyond the scope of this work. Nonetheless,
the high Mg abundances in the ejecta suggest a substantial amount was
synthesized in the explosion. In the nucleosynthesis models of
\citet{thielemann96}, significant amounts of Mg are only produced in
massive progenitors of $>25$ $\msun$.

\section{Conclusions}

While the northern region of G284 is dominated by shocked ISM, we have
shown that the ejecta-dominated western region is significantly
overabundant in Mg. Along with N49B in the LMC, this makes only the
second known remnant with Mg-rich ejecta. Models of nucleosynthesis in
CCSN progenitors produce significant amounts of Mg relative to other
products, like Si and Fe, only in the explosions of massive stars
($>25\ \msun$).

We find no statistically significant extended emission in the radial
profile for $r<7''$ from the binary J1018 at the center of G284, though
we do see a slight excess of photons a few arcseconds southeast of the
source. The spectral parameters of J1018 are about the same in the two
observations shifted by 0.25 in binary phase, but the flux difference
is significant, $\approx30$--$40\%$. Our binary evolution calculations
support the idea that the progenitor for this SN had an initial mass
of $>$ 25 $\msun$. Mass transfer prior to the explosion from the
progenitor to the current donor star increased that star's mass to
$\sim 30\ \msun$. Our simulations imply that the compact object in the
binary is a heavy, $\sim2\msun$, NS.

With its large size and low surface brightness, G284 would benefit
from a deep study with an X-ray telescope with both large effective
area and field of view. A long observation with the pn instrument on
{\it XMM-Newton} in Full Frame mode or the Soft X-ray Imager (SXI) on
the upcoming {\it Astro-H} satellite would be ideal for the purpose of
further study on the fainter sections of the ejecta.

\acknowledgments

We thank the anonymous referee for useful comments. We acknowledge
useful discussions with Steve Snowden on the {\it XMM-Newton} Extended
Source Analysis Software (ESAS), a part of the SAS software
distributed by the {\it XMM-Newton} Guest Observer Facility. We
acknowledge \emph{Chandra} Grant G03-14047.

\newpage
\clearpage

\begin{deluxetable}{lccccccccc}
\tablecolumns{10}
\tablewidth{0pc}
\tabletypesize{\footnotesize}
\tablecaption{X-ray Spectral Fits for G284}
\tablehead{
\colhead{Region} & $kT$ (keV) & Ne & Mg & Si (S) & Fe & $\tau_{i}$ (cm$^{-3}$ s) & Norm. & Flux & $\chi_\nu^{2}$ (d.o.f.)}

\startdata

North & 0.67$^{0.72}_{0.63}$ & 1.19$^{1.70}_{0.86}$ & 1.06$^{1.46}_{0.79}$ & 0.19$^{0.32}_{0.09}$ & 0.24$^{0.32}_{0.18}$ & 4.6$^{...}_{1.9}$ $\times 10^{12}$ & 9.6$^{11.7}_{7.7}$ $\times 10^{-4}$ & 3.1 & 1.96 (112)\\
West & 0.92$^{4.09}_{0.71}$ & 1.30$^{1.72}_{0.95}$ & 4.53$^{6.39}_{3.49}$ & 1.50$^{2.16}_{1.08}$ & 0.97$^{2.22}_{0.62}$ & 1.0$^{2.1}_{0.4}$ $\times 10^{11}$ & 1.6$^{2.6}_{0.7}$ $\times 10^{-4}$ & 2.5 & 1.40 (184)\\

\enddata

\tablecomments{Fits to the two regions of G284 shown in
  Figure~\ref{images}. Fits were performed using {\it XSPEC} with a
  \texttt {phabs*vpshock} model. Abundances are relative to
  solar. Subscripts and supercripts are lower and upper bounds,
  respectively, at the 90\% confidence level. Flux in units of
  10$^{-13}$~erg~s$^{-1}$~cm$^{-2}$ over range $0.5-2.5$~keV.}
\label{params}
\end{deluxetable}

\newpage
\clearpage

\end{document}